\documentstyle[aps,epsf,floats,twocolumn]{revtex}

\begin{document}
\draft

\title{Induced superconductivity distinguishes chaotic from
integrable billiards}
\author{J. A. Melsen, P. W. Brouwer, K. M. Frahm, and C. W. J. Beenakker}
\address{Instituut-Lorentz, University of Leiden\\
P.O. Box 9506, 2300 RA Leiden, The Netherlands \medskip\\
\parbox{14cm}{\rm
Random-matrix theory is used to show that the proximity to a superconductor
opens a gap in the excitation spectrum of an electron gas confined to a
billiard with a chaotic classical dynamics. In contrast, a gapless
spectrum is obtained for a non-chaotic rectangular billiard, and it is
argued that this is generic for integrable systems.\\
{\tt preprint: cond-mat/9604058}
}}
\maketitle
\narrowtext
The quantization of a system with a chaotic classical dynamics is the
fundamental problem of the field of ``quantum
chaos''\cite{Houches89,Gutzwillerbook}.
It is known that the statistics of the energy levels of a two-dimensional
confined region (a ``billiard'') is different if the dynamics is chaotic or
integrable\cite{Bohigas,Berry,AASA}: A chaotic billiard has Wigner-Dyson
statistics,
while an integrable billiard has Poisson statistics. The two types of
statistics are entirely different as
far as the level correlations are concerned\cite{Mehta}. However, the
mean level spacing is essentially the same: Particles of mass $m$ in a
billiard of area $A$ have
density of states $mA/2\pi\hbar^2$, regardless of whether their dynamics
is chaotic or not.

In the solid state, chaotic billiards have been realized in semiconductor
microstructures known as ``quantum \mbox{dots}'' \cite{Westervelt}.
These are confined regions in a two-dimensional electron gas, of
sufficiently small size that the electron motion remains ballistic and
phase-coherent on long time scales. (Long compared to the mean dwell time
$t_{\rm dwell}$
of an electron in the confined region, which itself is much longer than the
ergodic time $t_{\rm erg}$ in which an electron explores the available phase
space.)
A tunneling experiment
measures the density of states in the quantum dot, if its capacitance is
large enough that the Coulomb blockade can be ignored. As mentioned
above, this measurement does not distinguish chaotic from integrable
dynamics.

In this paper we show that the density of states becomes a probe for
quantum chaos if the electron gas is brought into contact with a
superconductor. We first consider a chaotic billiard. Using random-matrix
theory, we compute the density of states $\rho(E)$ near the Fermi level
$(E=0)$, and find that the coupling to a superconductor via a
tunnel barrier induces an energy gap $E_{\rm gap}$ of the order of the
Thouless energy $E_{\rm T}\simeq \hbar/t_{\rm dwell}$. More precisely,
\begin{equation}\label{eq:egap}
E_{\rm gap} = c N \Gamma \delta/2\pi,
\end{equation}
where $N$ is the number of transverse modes in the barrier, $\Gamma$ is the
tunnel probability per mode, $2\delta$ is the mean level spacing of the
isolated billiard, and $c$ is a number which is weakly dependent on $\Gamma$
($c$ decreases from 1 to 0.6 as $\Gamma$ increases from $0$ to $1$).
Eq.\ (\ref{eq:egap}) requires $1 \ll N\Gamma \ll \Delta/\delta$, where
$\Delta$ is the energy gap in the bulk of the superconductor. In this
limit $\rho(E)$ vanishes identically for $E \le E_{\rm gap}$. In contrast,
for a rectangular billiard we do not find an energy gap in which
$\rho=0$, but instead find that the density of states vanishes linearly
with energy near the Fermi level. We present a general argument that in
an integrable billiard $\rho$ has a power-law dependence on $E$ for small
$E$.

\begin{figure}
\epsfxsize=0.95\hsize
\epsffile{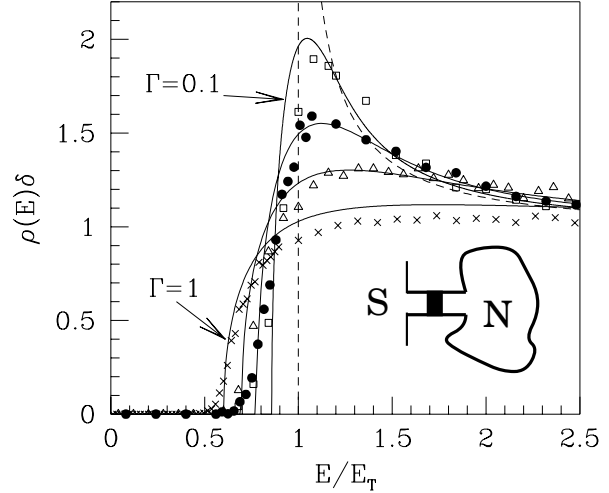}
\caption{\label{fig:Gfig1}
Density of states of a chaotic billiard coupled to a superconductor (inset),
for various coupling strengths.
The energy is in units of the Thouless energy $E_{\rm T}=N\Gamma\delta/2\pi$.
The solid curves are computed from Eqs.\ (\protect\ref{eq:onemoretrace}) and
(\protect\ref{eq:dysongamma}), for
$\Gamma = 1$, $0.5$, $0.25$, $0.1$.
The dashed curve is the asymptotic result (\protect\ref{eq:weakc})
for $\Gamma \ll 1$. The data points are a numerical solution of Eq.\
(\protect\ref{eq:HBdG}), averaged over $10^5$ matrices $H_0$ in the Gaussian
orthogonal ensemble ($M=400$, $N=80$). The deviation from the analytical
curves is mainly due to the finite dimensionality $M$ of $H_0$ in the
numerics.
}
\end{figure}

The system considered is shown schematically in the inset of
Fig.\ \ref{fig:Gfig1}. A confined region in a normal metal (N) is
connected to a superconductor (S) by a narrow lead containing a tunnel
barrier. The lead supports $N$ propagating modes at the Fermi energy.
Each mode may have a different tunnel probability $\Gamma_n$, but later
on we will take all $\Gamma_n$'s equal to $\Gamma$ for simplicity. The
proximity effects considered here require time-reversal symmetry, so
we assume zero magnetic field. (The case of broken time-reversal
symmetry has been studied previously \cite{Kosztin,Altland,Frahmetal}.)
The quasi-particle excitation spectrum of the system is discrete for
energies below $\Delta$. We are interested in the low-lying part of the
spectrum, consisting of (positive) excitation energies $E_n \ll \Delta$. We
assume that the Thouless energy $E_{\rm T} \equiv N\Gamma \delta/2\pi$ is also
much smaller than $\Delta$\cite{opposite}.

There are two methods to compute the spectrum in the regime
$E, E_{\rm T} \ll \Delta$. The first method is a scattering approach, which
leads to the determinant equation\cite{Carlo92}
\begin{equation} \label{eq:bound}
  \mbox{Det}[1+S^{\vphantom{*}}_0(E)S^*_0(-E)]=0.
\end{equation}
The $N \times N$ unitary matrix $S_0(E)$ is the scattering matrix of the
quantum dot plus tunnel barrier at an energy $E$ above the Fermi level.
Eq.\ (\ref{eq:bound}) is a convenient starting point for the case that
the quantum dot is an integrable billiard. For the chaotic case, we will
use an alternative --- but equivalent --- determinant equation involving an
effective Hamiltonian\cite{Frahmetal},
\begin{equation} \label{eq:HBdG}
  \mbox{Det}(E-H)=0,\; H = \left(\begin{array}{cc}
    H_0 & -\pi W W^{\rm T} \\
    - \pi W W^{\rm T} & -H_0^*
    \end{array}\right).
\end{equation}
The $M \times M$ Hermitian matrix $H_0$ is the Hamiltonian of the isolated
quantum dot. (The finite dimension $M$ is taken to infinity later on.)
Because of time-reversal symmetry, $H_0=H_0^*$.
The $M\times N$ coupling matrix $W$ has elements
\begin{eqnarray}
\nonumber
W_{mn} & = & \delta_{mn}\left(\frac{2M\delta}{\pi^2}\right)^{1/2}\! \!
\left(2\Gamma_n^{-1}-1+ 2\Gamma_n^{-1}\sqrt{1-\Gamma_n}\right)^{1/2},\\
\label{eq:coup_def}
&& m=1,2,\ldots M,\quad n=1,2,\ldots N.
\end{eqnarray}
The energy $\delta$ is one half the
mean level spacing of $H_0$, which equals the mean level spacing of
$H$ if $W=0$.

We now proceed to compute the density of states. We first consider the
case of a chaotic billiard. The Hamiltonian $H_0$ then has
the distribution of the Gaussian orthogonal ensemble \cite{Mehta},
\begin{equation} \label{eq:HGauss}
P(H_0) \propto
  \exp\left(-\case{1}{4} M \lambda^{-2}\, \mbox{Tr}\, H_0^2\right),\;
\lambda=2M\delta/\pi.
\end{equation}
We seek the density of states
\begin{equation}
\label{eq:resolvent}
  \rho(E) =
    -\pi^{-1} \mbox{Im}\, \mbox{Tr}\, \langle(E + i0^+ -H)^{-1}\rangle,
\end{equation}
where $\langle \cdots \rangle$ denotes an average over $H$ for fixed
$W$ and $H_0$ distributed according to Eq. (\ref{eq:HGauss}). The method
we use to evaluate this average is a perturbation expansion in $1/M$,
adapted from Refs.\cite{Pandey,brezinzee}.
Because of the block structure of $H$ [see Eq.\ (\ref{eq:HBdG})], the Green
function ${\cal G}(z) = \langle (z-H)^{-1}\rangle$ consists of four
$M\times M$ blocks ${\cal G}_{11}$, ${\cal G}_{12}$, ${\cal G}_{21}$, ${\cal
G}_{22}$.
By taking the trace of each block separately, one arrives at a $2\times 2$
matrix Green function
\begin{equation}
  {G} = \frac{\lambda}{M}
    \left( \begin{array}{cc} \mbox{Tr}\, {\cal G}_{11} & \mbox{Tr}\, {\cal
G}_{12} \\
           \mbox{Tr}\, {\cal G}_{21} & \mbox{Tr}\, {\cal G}_{22} \end{array}
\right).
\end{equation}
(We have multiplied by $\lambda/M=2\delta/\pi$ for later convenience.)
One more trace yields the density of states,
\begin{equation}\label{eq:onemoretrace}
\rho(E)=-\case{1}{2}\delta^{-1}\mbox{Im}\mbox{Tr}\, G(E+i0^+).
\end{equation}
To leading order in $1/M$, the matrix ${G}$ satisfies
\begin{equation}
\label{eq:pastur}
{G} = \frac{\lambda}{M}\sum_{n=1}^M \left(
\begin{array}{cc}
z-\lambda {G}_{11} & \pi w_n^2 +\lambda {G}_{12} \\
\pi w_n^2 +\lambda {G}_{21} & z-\lambda {G}_{22}
\end{array}\right)^{-1},
\end{equation}
where we have abbreviated $w_n^2=(W W^{\rm T})_{nn}$. Eq.\ (\ref{eq:pastur})
is a matrix-generalization of Pastur's equation \cite{Pastur}.
A unique solution
is obtained by demanding that ${G}$ goes to
$\lambda/z$ times the unit matrix as $|z| \rightarrow \infty$.

We now restrict ourselves to identical tunnel probabilities, $\Gamma_n \equiv
\Gamma$. For $M \gg N \gg 1/\Gamma$
Eq.\ (\ref{eq:pastur}) simplifies to
\begin{mathletters}
\label{eq:dysongamma}
\begin{eqnarray}
& & N{G}_{11}\delta  = \pi z {G}_{12}(-{G}_{12}+1-2/\Gamma),\\
& & {G}_{22}={G}_{11},\ {G}_{21}={G}_{12},\ {G}_{12}^2 = 1+{G}_{11}^2.
\end{eqnarray}
\end{mathletters}%
This set of equations can be solved analytically \cite{Josephson}. The result
is that $\rho(E)=0$ for $E\leq E_{\rm gap}$, where $E_{\rm gap}$ is determined
by
\begin{eqnarray}
& &  \frac{k^6-k^4}{(1-k)^6} x^6 - \frac{3k^4-20k^2+16}{(1-k)^4} x^4 +
    \frac{3 k^2+8}{(1-k)^2} x^2 = 1,\nonumber \\
& & x=E_{\rm gap}/E_{\rm T},\; k=1-2/\Gamma.
\end{eqnarray}
The solution of this gap equation is the result (\ref{eq:egap})
announced in the introduction. The
complete analytical solution of Eq.\ (\ref{eq:dysongamma}) is omitted here
for lack of space. In Fig.\ \ref{fig:Gfig1} we plot the resulting density of
states. In the limit $\Gamma=1$ of ideal coupling it is given by
\begin{mathletters}
\label{eq:dos}
\begin{eqnarray}
  & & \rho(E) =
   {E_{\rm T}\sqrt{3} \over 6 E \delta} [ Q_{+}(E/E_{\rm T}) -
     Q_{-}(E/E_{\rm T})], \\
   & & Q_{\pm}(x) = \left[8 - 36 x^2 \pm 3 x \sqrt{3 x^4 + 132 x^2 -
48}\right]^{1/3}, \\
    & & \lefteqn{E> E_{\rm gap} = 2 E_{\rm T}\,
       \gamma^{5/2} \approx 0.6\, E_{\rm T},} \ \ \  \label{eq:gapenergy}
\end{eqnarray}
\end{mathletters}%
where $\gamma = \frac{1}{2}(\sqrt{5} - 1)$ is the golden number. In the
opposite limit $\Gamma \ll 1$  of weak coupling we find
\begin{eqnarray} \label{eq:weakc}
   \rho(E) &=&  E{\delta}^{-1}
  ({E^2 - E_{\rm T}^2})^{-1/2}, \ E >E_{\rm gap}= E_{\rm T}.
\end{eqnarray}

To check the validity of the perturbation theory, we have computed $\rho(E)$
numerically from Eq.\ (\ref{eq:HBdG}) by generating a
large number of random matrices $H_0$ in the Gaussian orthogonal ensemble.
The numerical results (data points in Fig.\ \ref{fig:Gfig1}) are consistent
with Eq.\ (\ref{eq:dysongamma}), given the finite dimensionality of $H_0$
in the numerics.

\begin{figure}
\epsfxsize=0.95\hsize
\epsffile{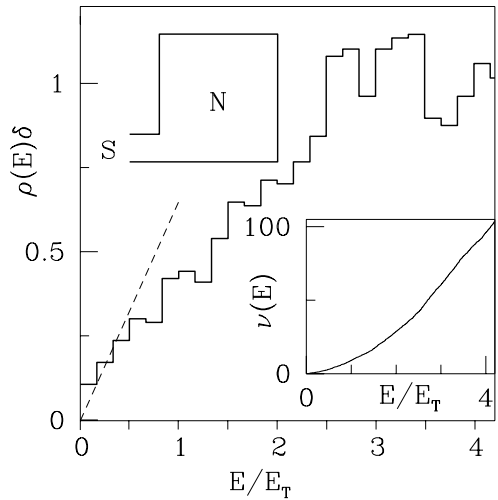}
\caption{\label{fig:sqdens}Histogram: density of states for
a rectangular billiard (shown to scale in the upper left inset), calculated
numerically from Eq.\ (\protect\ref{eq:bound}). Dashed curve:
Bohr-Sommerfeld approximation (\protect\ref{eq:BSsmallE}). The
lower right inset shows the
integrated density of states, which is the quantity following directly
from the numerical computation. The energy $E_{\rm T} = N\delta/2\pi$,
with $N=200$ modes in the lead to the superconductor.
}
\end{figure}

We now turn to a non-chaotic, rectangular billiard. A lead perpendicular
to one of the sides of the rectangle connects it to a superconductor.
(The billiard is drawn to scale in the upper left inset of Fig.\
\ref{fig:sqdens}).
There is no tunnel barrier in the
lead. The scattering matrix $S_0(E)$ is computed by matching wave
functions in the rectangle to transverse modes in the lead.
The density of states then follows from Eq.\ (\ref{eq:bound}). To improve the
statistics, we averaged over 16 rectangles with small differences in shape
but the same area $A$ (and hence the same $\delta=\pi \hbar^2/mA$). The
number of modes in the lead (width $W$) was fixed at $N=m v_{\rm F}
W/\pi\hbar=200$ (where $v_{\rm F}$ is the Fermi velocity).
In the lower right inset of Fig.\ \ref{fig:sqdens} we show the
integrated density of states $\nu(E)=\int_0^E dE'\rho(E')$, which is the
quantity
following directly from the numerical computation. The density of states
$\rho(E)$ itself is shown in the main plot.

We have also computed the Bohr-Sommerfeld approximation to the density
of states,
\begin{equation}\label{eq:BS}
\rho_{\rm BS}(E) = N\int_0^{\infty}\!\!\! d s\, P(s)
\sum_{n=0}^{\infty}\delta\left(E-(n+\case{1}{2})\pi \hbar v_{\rm F}/s\right).
\end{equation}
Here $P(s)$ is the classical
probability that an electron entering the billiard will exit after a
path length $s$. Eq.\ (\ref{eq:BS}) is the Bohr-Sommerfeld quantization
rule for the classical periodic motion with path length $2s$ and phase
increment per period of $2E s/\hbar v_{\rm F}-\pi$. The periodic motion
is the result of Andreev reflection at the interface with the
superconductor, which causes the electron to retrace its path as a hole.
The phase increment consists of a part $2E s/\hbar v_{\rm F}$ because of
the energy difference $2 E$ between electron and hole, plus a phase
shift of $-\pi$ from two Andreev reflections.
For $s \rightarrow \infty$ we find $P(s) \rightarrow 8(A/W)^2 s^{-3}$,
which implies a linear $E$-dependence of the density of
states near the Fermi-level,
\begin{equation}
\rho_{\rm BS}(E) \rightarrow \frac{4E}{N\delta^2}=\frac{2E}{\pi E_{\rm
T}\delta},\hspace{1cm}  E\rightarrow 0.
\label{eq:BSsmallE}
\end{equation}
In Fig.\ \ref{fig:sqdens} we see that
the exact
quantum mechanical density of states also has (approximately) a linear
$E$-dependence near $E=0$, but with a smaller slope than the semi-classical
Bohr-Sommerfeld approximation.

We argue that the absence of an excitation gap found in the rectangular
billiard is generic for the whole class of integrable billiards.
Our argument is based on the Bohr-Sommerfeld approximation.
It is known\cite{Bauer,Baranger} that an integrable billiard has
a power-law distribution
of path lengths, $P(s) \rightarrow s^{-p}$ for $s\rightarrow \infty$. Eq.\
(\ref{eq:BS}) then implies a power-law density of states, $\rho(E) \propto
E^{p-2}$
for $E\rightarrow 0$.

To conclude, we have shown that the presence of an excitation gap
in a billiard connected to a superconductor
is a signature of quantum chaos, which is special in
two respects: It appears in the spectral density rather than in a spectral
correlator, and it manifests itself on the macroscopic energy scale
of the Thouless energy
rather than on the microscopic scale of the level spacing. Both these
characteristics are favorable for experimental observation. Our
theoretical results are rigorous for a chaotic billiard and for an
integrable rectangular billiard. We have presented an argument that the
results for the rectangle are generic for the whole class of integrable
billiards, based on the semi-classical Bohr-Sommerfeld approximation.
It remains a challenge to develop a rigorous general theory for the
integrable case.

This work was supported by the Dutch Science Foundation NWO/FOM and by the
Human Capital and Mobility program of the European Community.


\begin{references}
\bibitem{opposite} The opposite regime $E_{\rm T} \gg \Delta$ has a trivial
discrete spectrum, consisting of $N$ states with energies $E_n$ just below
$\Delta$.
\end{references}
\end{document}